\documentclass{ifacconf}

\usepackage{graphicx}      
\usepackage{natbib}        
\usepackage{amsmath} 
\usepackage{amssymb}  
\usepackage{xcolor}
\usepackage{natbib}        
\usepackage{amsmath,amssymb}
\usepackage{xcolor}
\usepackage{algorithm}
\usepackage{algpseudocode}
\usepackage{multicol}
\usepackage{lipsum}
\usepackage{float}
\usepackage{multirow}
\usepackage{subfigure}
\DeclareMathOperator*{\argmin}{arg\,min}

\begin{document}
\begin{frontmatter}

\title{Real-Time Optimal Design of Experiment for Parameter Identification of Li-Ion Cell Electrochemical Model} 

\author[1]{Ian Mikesell} 
\author[1]{Samuel Filgueira da Silva} 
\author[1]{Mehmet Fatih Ozkan}
\author[1]{Faissal El Idrissi}
\author[1]{Prashanth Ramesh}
\author[1]{Marcello Canova}

\address[1]{Department of Mechanical and Aerospace Engineering, Center for Automotive Research, The Ohio State University, Columbus, OH, USA.}

\begin{abstract}                
 Accurately identifying the parameters of electrochemical models of li-ion battery (LiB) cells is a critical task for enhancing the fidelity and predictive ability. Traditional parameter identification methods often require extensive data collection experiments and lack adaptability in dynamic environments. This paper describes a Reinforcement Learning (RL) based approach that dynamically tailors the current profile applied to a LiB cell to optimize the parameters identifiability of the electrochemical model. The proposed framework is implemented in real-time using a Hardware-in-the-Loop (HIL) setup, which serves as a reliable testbed for evaluating the RL-based design strategy. The HIL validation confirms that the RL-based experimental design outperforms conventional test protocols used for parameter identification in terms of both reducing the modeling errors on a verification test and minimizing the duration of the experiment used for parameter identification.
\end{abstract}


\end{frontmatter}

\section{Introduction}
Lithium-ion batteries (LiBs) have become indispensable across diverse industrial sectors, being the energy storage solution of choice for applications spanning consumer electronics, electric vehicles, and grid-level energy storage. Accurate modeling and testing of LiBs are critical for optimizing their performance, ensuring safety, and extending their operational life. Historically, research in this domain has prioritized the development of physics-based electrochemical models and estimation algorithms. However, recent studies underscore the importance of parameter identification and model calibration, including the design of appropriate experiments, data collection and post-processing to facilitate such tasks \citep{anderson2022}.\par 
Among emerging methodologies in this field, Optimal Experimental Design (OED) has gained prominence for its ability to enhance parameter identifiability by maximizing Fisher Information (FI) for specific parameters over a defined time horizon \citep{huang2023excitation}. OED provides a systematic framework to identify operating conditions and excitation inputs that yield the most informative data for parameter calibration. As shown by \citep{Lopez2016} only a small number of parameters can be identified from conventional characterization tests such as constant discharge experiments while parameters like the rate constant in the anode are left unidentifiable. OED addresses these limitations by tailoring input profiles to ensure maximum informativeness for parameter estimation.

Despite its advantages, conventional OED methods face several challenges. For instance, input profiles are often selected from predefined libraries, which may exclude truly optimal designs. Convex programming approaches, as described in \citep{park2018optimal}, optimize FI but are limited by the constraints of the predefined library, potentially overlooking profiles that excite the battery in conditions most conducive to parameter identifiability. Iterative optimization schemes, such as those proposed by \citep{Pozzi_OED}, risk failing to converge to an optimal design due to computational limits. The widely used D-optimal design approach maximizes the determinant of the FI matrix but assumes linearity \citep{MATHIEU2017}, a limitation when dealing with the nonlinear behavior of high-fidelity electrochemical models \citep{Chhajer2024}.

To address these gaps, reinforcement learning (RL) has emerged as a promising tool for designing adaptive excitation policies that optimize data collection in real-time \citep{huang2023reinforcement}. By dynamically tailoring input profiles during experimentation, RL-based OED can improve parameter identifiability while reducing the computational burden and duration of the identification process. However, most studies to date have been confined to simulations, limiting their applicability to real-world scenarios \citep{LibErrors}.

\begin{figure*}[ht!]
  \centering
  \includegraphics[width=0.9\textwidth]{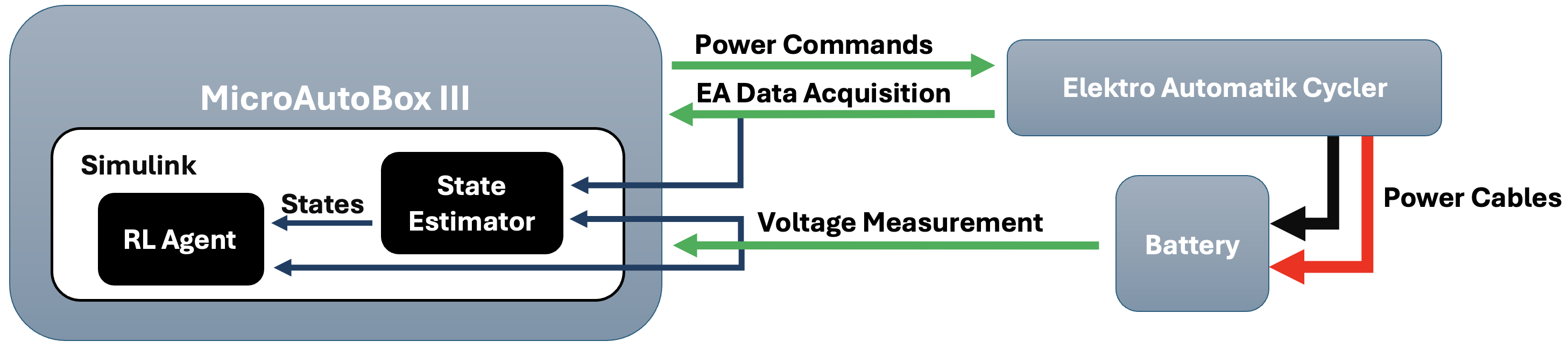}    %
\caption{Schematic of HIL experimental setup.} 
\label{fig:HIL Setup Schematic}
\end{figure*}

This paper builds on prior work by integrating an RL framework into a hardware-in-the-loop (HIL) environment to perform online OED. To illustrate the methodology, the paper focuses on the experimental procedure to estimate a rate constant parameter in a LiB electrochemical model. Unlike conventional characterization methods, such as constant-current discharges or relaxation current interrupt discharge (RCID) tests, the proposed approach dynamically adapts the input excitation to maximize the informativeness of collected data. This innovation enables improved parameter identifiability while reducing the duration and computational burden of the identification process.

The novelty of this work lies in leveraging HIL validation to bridge the gap between simulation and real-world experimentation. While simulations offer valuable insights, they often fail to fully account for real-world factors such as manufacturing inconsistencies, degradation mechanisms, or measurement errors \citep{LibErrors}. By implementing OED in a physical HIL environment, this study provides a robust evaluation platform for LiB parameter identification methods, supporting applications ranging from state of charge (SOC) estimation \citep{HIL_testingSOC} to battery management system (BMS) verification \citep{HIL_testingBMS}.

A reduced-order enhanced single-particle model (ESPM) is used to guide the OED process, serving as a benchmark against which experimental results are evaluated \citep{seals2022physics}. By comparing RL-based OED with conventional testing protocols, this study demonstrates the potential for significant improvements in both parameter identifiability and experimental efficiency.

\section{Experimental Setup}
The HIL environment for online OED validation is configured using MathWorks and dSPACE tools in conjunction with an Elektro Automatik (EA) Cycler.  More specifically, Simulink is used to initialize the test procedure which is ran through dSPACE ControlDesk.  The analog signals defining the set and measured voltage and current values are used to control charge and discharge current of the cycler.  A schematic of this flow of information can be seen in Fig. \ref{fig:HIL Setup Schematic}, where \emph{Power Commands} represents the voltage and current commands delivered to the cell cycler, \emph{EA Data Acquisition} includes the measured current values, \emph{Power Cable} indicates the current supply to the battery, and \emph{Voltage Measurement} is applied directly at the battery tabs to increase resolution and minimize sources of error.  

The measured voltage and current is fed into a data-driven State Estimator in Simulink. The estimator provides the normalized lithium surface concentration  in the cathode $\overline{c}_{se,p}$, which is required by the RL agent along with the measured terminal voltage. 

\section{Problem Formulation}
Consider a nonlinear, single-input, single-output discrete-time system:
\begin{equation}
    x_{k+1} = f_k(x_{k},u_{k}, \Theta)
\end{equation}
\begin{equation}
    y_k = g_k(x_k, u_k, \Theta)
\end{equation}
where \( x_k \) denotes the states, \( u_k \) is the input, \( y_k \) the voltage output, and \(\Theta =\{\theta_1,\theta_{2},\theta_{3},...,\theta_{m}\}\) the entire set of model parameters, including known parameters and parameters that need to be identified. 

Without loss of generality, consider the case of single parameter identification, $\Theta = \theta$. The objective of OED is to define the input sequence \( u_1, u_2, \dots, u_N \) over a finite time horizon \( N \) that maximizes the information content of the parameter $\theta$ on the output \( y_k \), which is achieved by optimizing the FI \citep{scharf1993geometry}:
\begin{equation}
\label{eq:FI}
    FI = \frac{1}{\sigma_y^2} \sum_{k=1}^{N} \left( \frac{\partial y_k}{\partial \theta} \right)^2
\end{equation}
where \( \sigma_y^2 \) denotes the variance of the output measurement noise. In the context of LiB cell model calibration, maximizing FI implies optimizing the input current profile to enhance the sensitivity of the voltage output to the parameter \citep{huan22vppc}. The OED problem can be formulated as:
\begin{equation}
\begin{aligned}
\operatorname*{max}_{u_1, u_2, \dots, u_N} \sum_{k=1}^{N} \left( \frac{\partial y_k}{\partial \theta} \right)^2 \\
\text{s.t.} \quad x_{k+1} = f_k(x_k, u_k, \Theta) \\
\quad y_k = g_k(x_k, u_k, \Theta)
\end{aligned}
\end{equation}
The OED problem requires the physics-based LiB model to compute the states, output and sensitivity of the voltage output \( y_k \) to the target parameter \( \theta \).

\subsection{Overview of Li-ion Cell Model Equation}
This study leverages the electrochemical model of a 30 Ah NMC-graphite cell for automotive applications. A reduced-order electrochemical equivalent circuit model (E-ECM) was previously developed and compared against a high-fidelity physics-based model of the same cell \citep{seals2022physics}. The benefit to utilizing the E-ECM compared to other electrochemical models is that it reduces the computation time for simulation. 

A summary of the governing equations for the E-ECM is provided below, while a more in-depth study of the framework can be found in \citep{seals2022physics}. The terminal voltage $V$ of the E-ECM comprises seven terms:
\begin{equation} \label{eq:1}
\begin{split}
V(t) = U_p(c_{se,p},t) - U_n(c_{se,n},t) - \eta_p(c_{se,p},t) \\ + \eta_n(c_{se,n},t) + \phi_{diff}(t) + \phi_{ion}(t) - I(t)R_c
 \end{split}
\end{equation}
The first two terms, $U_p$ and $U_n$, are the half-cell open circuit potentials of the cathode and anode, which are nonlinear functions of the lithium surface concentration, $c_{se,i}, i=p,n$, in each electrode. The terms, $\eta_{i}$, define the kinetic overpotential of each electrode at the solid-electrolyte interface, computed through a linearized Butler-Volmer reaction kinetic equation. The overpotentials in the electrolyte due to diffusion and ionic conductivity are represented by the terms $\phi_{diff}$ and $\phi_{ion}$, respectively. The $R_c$ represents the contact resistance at the current collectors. 

Last, the surface concentration dynamics \eqref{eq:7} is predicted by a linear system, obtained via Pade approximation of the diffusion partial differential equations (PDEs) describing the transport of lithium in the solid phase \citep{seals2022physics}:
\textcolor{black}{\begin{equation} \label{eq:7}
\begin{split}
c_{se,i}(s) = c_{se,i0} + (G_b(s) + G_d(s))\frac{-R_i}{3F\epsilon_{am,i}L_iA} I(s)
 \end{split}
\end{equation}}
\begin{equation} \label{eq:8}
G_b(s) = \frac{\frac{2}{7}\frac{R_i}{D_i}s + \frac{3}{R_i}}{\frac{1}{35}\frac{R_i^2}{D_i}s^2 + s}, \quad G_d(s) = \frac{\frac{5D_i}{R_i}}{s + \frac{7}{R_i}}
\end{equation}
where $G_b(s)$ represents the bulk concentration dynamics and $G_d(s)$ the diffusion dynamics, $R_i$ is the particle radius and $D_i$ denotes the diffusion coefficient, a parameter assumed known.

\subsection{Calculation of Parameter Sensitivity}
Because the li-ion cell model is physics-based, hence described fully by analytical equations, the sensitivity of the terminal voltage in \eqref{eq:1} to a given parameter is assessed by symbolically computing the first-order partial derivative with respect to that parameter. In this paper, the parameter considered as a case study to illustrate the OED procedure is the anode rate constant ($k_n$). This parameter is critical for accurately capturing the dynamic behavior of the electrochemical model \citep{dangwal2021parameter}.

The sensitivity of the cell voltage to the parameter $k_n$ is given by:
\textcolor{black}{\begin{equation} \label{eq:2}
\begin{split}
\frac{\partial V(t)}{\partial k_i} = \frac{\partial U_p(t)}{\partial k_i} - \frac{\partial \eta_p(t)}{\partial k_i} - \frac{\partial U_n(t)}{\partial k_i} +  \frac{\partial \eta_n(t)}{\partial k_i} - \\ \frac{\partial \phi_e(t)}{\partial k_i}  - \frac{\partial R_c(t)}{\partial k_i}
 \end{split}
\end{equation}}
which, after simplification, becomes:
\textcolor{black}{\begin{equation} \label{eq:3}
\ \ \frac{\partial V(t)}{\partial k_n} =   \frac{\partial \eta_n(t)}{\partial i_{0n}} \frac{\partial i_{0n}(t)}{\partial k_n} 
\end{equation}}
where $\eta_{i}$ represents the kinetic overpotential and $i_{0i}$ denotes the exchange current density. Interested readers are encouraged to see \citep{seals2022physics} for the definitions of $\eta_{i}$ and $i_{0i}$. Substituting the analytical expressions for each PDE $\left(\frac{\partial \eta_n(t)}{\partial i_{0n}}, ~ \frac{\partial i_{0n}(t)}{\partial k_n} \right)$, the sensitivity to $k_n$ can be expressed as:
\begin{equation} \label{eq:3_1}
\begin{split}
\frac{\partial V(t)}{\partial k_n} = \left( \frac{\overline{R} T_0 \left(J_n I(t)\right)}{F i_{0,n}^2} \exp\left(\left(\frac{1}{T_{\text{ref}}} - \frac{1}{T(t)}\right) \frac{E_{i0n}}{\overline{R}}\right) \times \right. \\
\left. F \sqrt{c_{se,n} \left(c_{\text{max},n} - c_{se,n}\right)c_e} \right) 
\end{split}
\end{equation}
where $J_n$ represents the intercalation current density at the anode, $E_{o,n}$ denotes the activation energy associated with the anode, $c_{max,n}$ is the maximum surface concentration in the anode, $c_e$ indicates the electrolyte concentration, and $T_{ref}$ specifies the reference temperature.

\section{Optimal Experimental Design with Reinforcement Learning}
This work employs the Twin Delayed Deep Deterministic Policy Gradient (TD3) algorithm, an advanced actor-critic Deep RL (DRL) method, to optimize input excitation in LiBs. TD3, an enhancement of the Deep Deterministic Policy Gradient (DDPG) algorithm, is chosen for its improved stability and effectiveness in continuous action spaces, making it ideal for this application \citep{fujimoto2018addressing}.

The problem of generating optimal input excitation is modeled as a Markov Decision Process (MDP), where the states ($s$) are defined by measurable or estimable quantities. In this work, the measured cell terminal voltage ($V$) and the estimated normalized lithium surface concentration in the cathode ($\overline{c}_{se,p}$) are assumed as the states for the model. The goal is to determine an optimal control policy, $\pi(s|a)$, that maximizes the reward function ($r$), mapping states $s$ to action ($a$), where $a$ represents the electrical current. RL learns this policy by exploring the state-action space and solving the Bellman Equation. The Bellman Equation evaluates the value of a state $s$ by combining immediate rewards with discounted future rewards, guiding the optimization process. The Q-function is defined as:
\begin{equation}
Q\left(s \mid a\right) = \max_{\pi(s \mid a)} \mathbb{E}\left[r_t + \sum_{k=1}^{\infty} \gamma^k r_{t+k} \mid s_{t+1}, a_{t+1}\right],
\end{equation}
evaluates the joint desirability of a state-action pair under a given policy $\pi(s|a)$, where $\gamma$ is the discount factor and $r_{t+k}$ represents the reward at time $t+k$.

The TD3's actor-critic structure consists of two networks: the actor network, which learns the policy $\pi(s|a)$, and the critic network, which estimates Q-values. The reward function is designed with two components. The first term maximizes the normalized FI for the selected parameter, while the second term penalizes any constraint violation on the cell voltage. The single-step reward is given as:
\begin{equation}
r_k =
\begin{cases}
    \left(\frac{\partial V_k}{\partial k_n}k_n\right)^2, & \text{normal condition} \\
    M, & \text{constraint violation}.
\end{cases}
\end{equation}\par
Each training episode begins at 100\% SOC and either concludes after 1800 steps or terminates early if constraints are violated. A step size of 1 second is applied, with training conducted over 1,000 to 5,000 episodes. 

The TD3 algorithm is trained using MATLAB's RL toolbox. Table \ref{table:setting} details the operating conditions of the simulated battery system and penalties for constraint violations. 
\begin{table}[h!]
\centering
\caption{TD3 Model Parameters and Settings}
\label{table:setting}
\begin{tabular}{|l|l|}
\hline
\textbf{Parameter} & \textbf{Value} \\ \hline
Current Input Range & \([-2\text{C} , 5\text{C}]\) \\ \hline
Voltage Output Range & \([2.8 \text{ V}, 4.2 \text{ V}]\) \\ \hline
Penalty for Constraint Violation \(M\) & -5 \\ \hline
Simulation Length  and Time step & 1800 s and 1 s\\ \hline
\end{tabular}
\label{table:model_params}
\end{table}
\section{Data-Driven State Estimator}

A data-driven identification approach is used to create an estimator of the normalized lithium surface concentration in the cathode $\overline{c}_{se,p}$, which is required as a state by the RL agent. This approach is warranted by the need to achieve a generally applicable procedure that could be applied even if the plant model is unknown. The method used to synthesize the data-driven model is based on a sparse identification of nonlinear dynamics with control (SINDYc) \citep{BRUNTON2016710}, which constructs a dynamic model as follows:
\begin{equation} \label{eq:21}
\mathbf{{\overline{c}_{se,p}}}[k+1] = \Theta(\mathbf{\overline{c}_{se,p}}[k],\mathbf{V}[k],\mathbf{I}[k])\xi
\end{equation}
\noindent where $\Theta \in \mathbb{R}^{n \times m}$ represents the basis functions of the predicted model, and $\xi \in \mathbb{R}^{m}$ is the sparse coefficient vector. The $\xi$ vector is determined by a sequentially thresholded ridge (STRidge) regression technique (\ref{eq:24}), where $\lambda_1$ and $\lambda_2$ represent the regularization parameter and sparsification knob, respectively. The selection of these hyperparameters and the library functions $\Theta$ is achieved through a hybrid sparse learning approach, as outlined in \citep{ga_sindy}.
\begin{equation}  \label{eq:24}
\begin{split}
    \hat{\xi} = \arg \min_{\xi} \left\| \mathbf{\overline{c}_{se,p}}[k+1] -  \Theta(\mathbf{\overline{c}_{se,p}}[k],\mathbf{V}[k],\mathbf{I}[k])\xi \right\|_{2} \\ + \lambda_{1} \left\| \xi \right\|_{2} \\ \quad \text{s.t.:} \quad |\hat{\xi}_i| < \lambda_2
\end{split}
\end{equation}

The current profile used in the training dataset is shown in Fig. \ref{fig:SOC_1}. This profile was generated by simulating the battery cell under various current cycles. To avoid overdischarge, a constant charging current was applied between each cycle. Validation of the model was performed by using current-voltage data corresponding to the regulatory UDDS, HWFET, US06 and WLTC driving cycles. Figure \ref{fig:SOC_2} illustrates the model's performance on the training dataset, achieving an R$^2$ value of 0.9987 and a mean squared error (MSE) of of 1.16 $\times 10^{-4}$. Similarly, as shown in Figure \ref{fig:SOC_3}, the model effectively captures the $\overline{c}_{se,p}$ dynamics during validation, with an R$^2$ value of 0.9951 and an MSE of 4.28 $\times 10^{-4}$. 

As detailed in the following section, the data-driven estimator closes the loop in the HIL framework by providing the state estimate $\overline{c}_{se,p}$ that is fed back to the RL agent.

\begin{figure}[h!] \label{fig:SOC_1}
    \begin{center}
       \includegraphics[angle=0,scale=0.63]{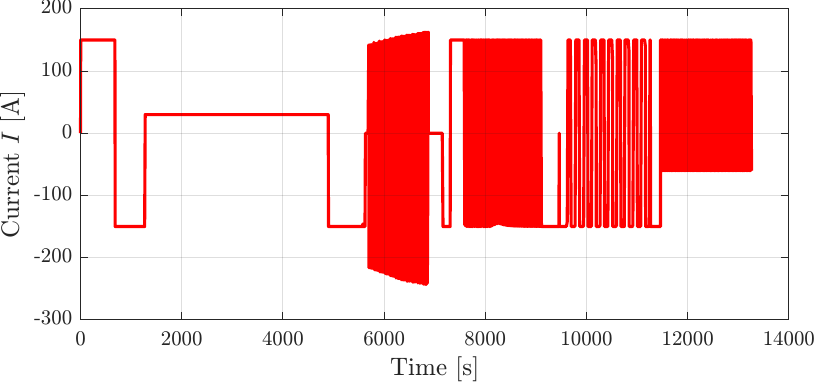}
    \caption{Current profile for training}
    \label{fig:SOC_1}
    \end{center}
\end{figure}

\begin{figure}[h!] \label{fig:SOC_2}
    \begin{center}
       \includegraphics[angle=0,scale=0.63]{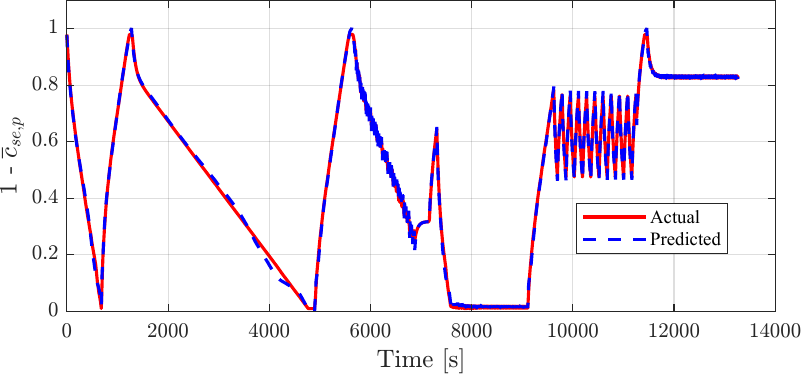}
    \caption{Battery model state $\overline{c}_{se,p}$ prediction for training set}
    \label{fig:SOC_2}
    \end{center}
\end{figure}

\begin{figure}[h!] \label{fig:SOC_3}
    \begin{center}
       \includegraphics[angle=0,scale=0.63]{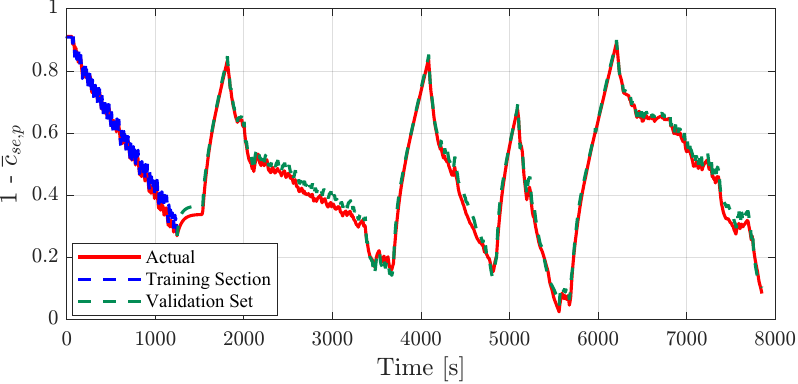}
    \caption{Battery model state $\overline{c}_{se,p}$ prediction under a section of training and validation sets}
    \label{fig:SOC_3}
    \end{center}
\end{figure}

\section{Implementation, Results and Analysis}
As mentioned above, the focus of this study is to demonstrate the process of OED for the parameter \( k_n \), facilitated through an RL agent that makes decisions in real-time through a HIL framework with a battery cell in the loop. 

\begin{figure}[h!]
\begin{center}
\includegraphics[width=8.4cm]{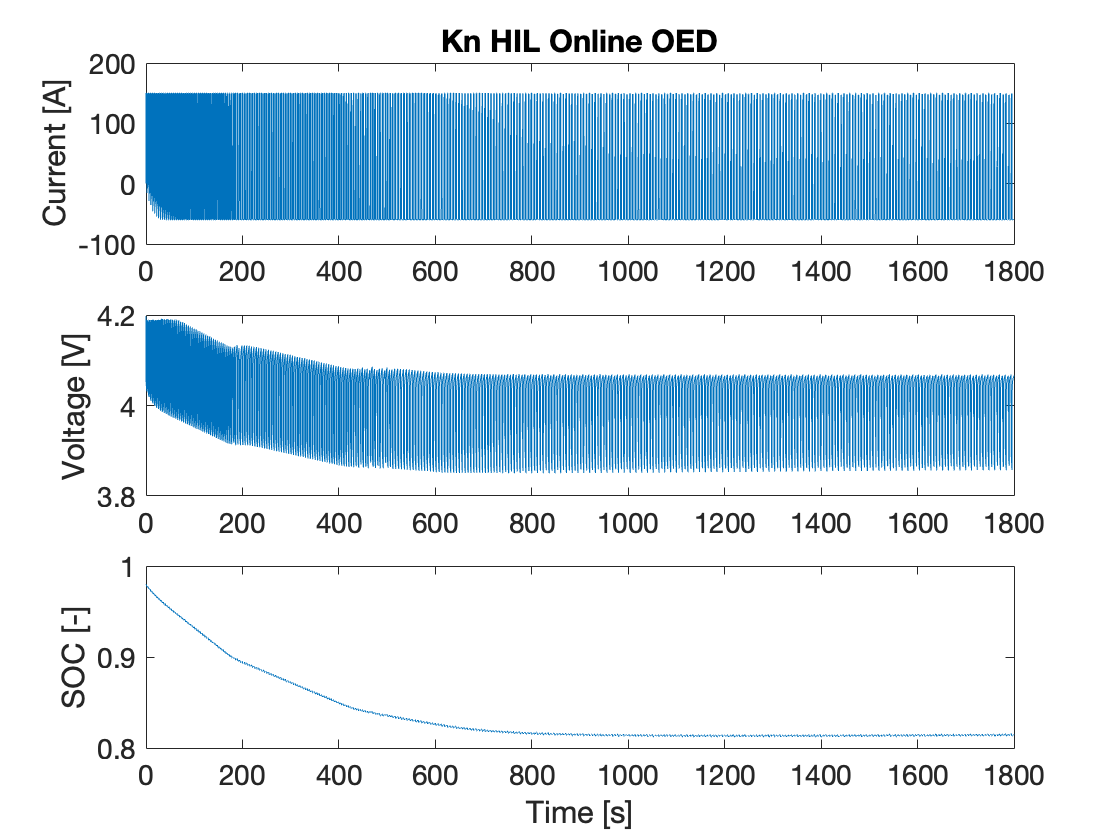}    
\caption{RL-HIL experimental data for $k_n$ identification}
\label{fig:Kn experimental results}
\end{center}
\end{figure}

Figure \ref{fig:Kn experimental results} illustrates the results of the experimental test, including the input current, battery SOC (calculated via Coulomb counting) and measured cell terminal voltage, which are pivotal for understanding the system's behavior under the designed conditions.

In the high SOC region, the lithium concentration within the anode is significant, nearing its upper boundary as the SOC approaches 100\%. This results in increased voltage sensitivity to variations in \( k_n \), making this region highly informative for calibrating \( k_n \). The RL agent identifies this high-sensitivity operating condition from and guides the battery to operate predominantly in this region. The resulting excitation, as evident from the applied current and corresponding voltage fluctuations in the figure, is designed to extract the maximum information relevant to \( k_n \).

\subsection{Parameter Identification Analysis}
After collecting the data using the HIL setup, an offline parameter identification process is conducted to compare the current profile generated via the RL-based OED to a set of conventional test cycles commonly used in battery testing, including 1C and 5C constant current discharging tests and RCID \citep{hu2011electro}.

The problem of parameter identification is addressed as a model-free constrained optimization, where a Bayesian optimization method \citep{pi2024parameter} is applied to minimize the squared error between experimental voltage measurements and the E-ECM voltage output starting from an arbitrary initial value of the parameter $k_n$. The optimization problem is expressed as follows:
\begin{equation}
\theta_i^\ast = \argmin_{\theta \in \Theta}\sum_{k=0}^{N-1} \left( V_{exp} \left(I_k \right) - {V}_k \left( {\theta}_i, I_k \right) \right)^2
\end{equation}
The constraint set $\Theta$ for \(k_n\) is defined as $[2\times 10^{-11}, 1\times 10^{-8}]$ and the evenly distributed ten initial guesses for \(k_n\) from this set are considered in the optimization process.

To validate the effectiveness of the identification process, a drive cycle from \citep{pi2024parameter} is considered as the verification test. This involves using the identified parameter \(k_n\) from each experimental design to predict the battery's voltage response under a realistic drive cycle. The predicted voltage is then compared against the experimental voltage measurements, and the median of the prediction errors across the drive cycle is calculated. 

\begin{figure}[h]
\begin{center}
\includegraphics[width=8.4cm]{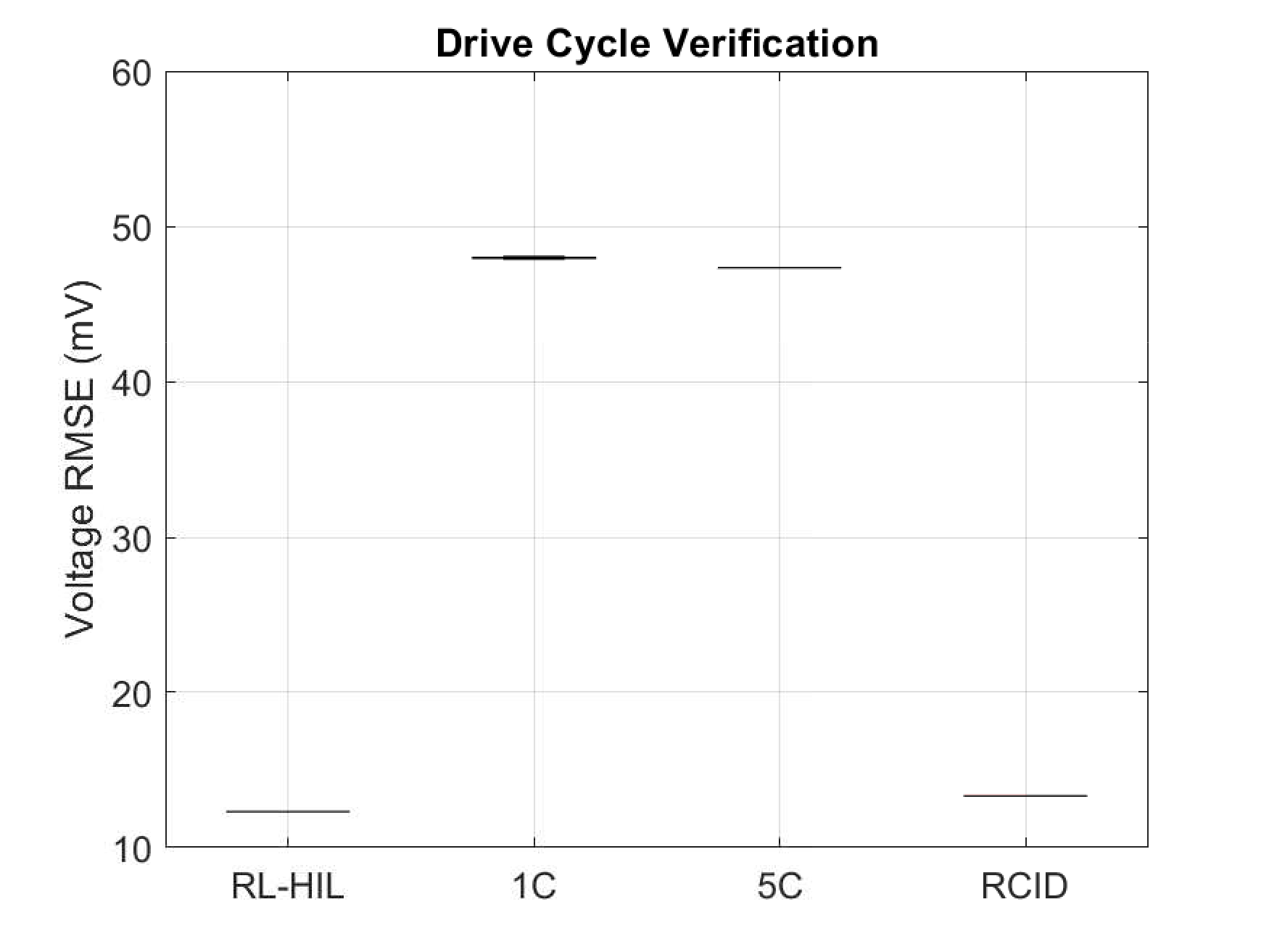}    
\caption{Drive cycle verification of the calibrated model with identified $k_n$ using RL-HIL experiment and 1C, 5C and RCID conventional tests}
\label{fig:drivecylce_verification}
\end{center}
\end{figure}

To illustrate the results, Fig. \ref{fig:drivecylce_verification} shows a box plot computing the prediction errors across all tests, providing an overview of the variability and consistency in the prediction errors for each experimental design. The variation in errors is minimal across all tests, indicating that the Bayesian optimization process consistently converges to the same value regardless of the initial guesses. This consistency demonstrates the robustness of the Bayesian optimization method in parameter identification. The median values of these verification errors, as derived from the box plot, are listed in Table \ref{table:2} for a more detailed comparison.
\begin{table*}[h!]
\centering
\caption{Parameter identification results for various experimental designs, showing the FI, drive cycle verification error (median), and experiment length for each test}
\label{table:2}
\begin{tabular}{cccc} 
\hline
{\textbf{Current profile}} & {\textbf{FI [V$^2$]}} & {\textbf{Drive Cycle Verification Error (Median) [mV]}} &{\textbf{Experiment Length [s]}} \\ 
\hline
1C Discharge& 0.01 & 47.98 & 3781 \\
5C Discharge& 0.05 & 47.31 & 701 \\
RCID& 0.42 &  13.33 & 103333  \\
\textbf{RL-HIL}          & \textbf{0.08} &  \textbf{12.32} & \textbf{1800} \\
\hline
\end{tabular}
\end{table*}

The results in Table \ref{table:2} highlight the effectiveness of different current profiles for parameter identification, with the profile obtained from the RL-based OED demonstrating a strong balance between accuracy and test duration. The lowest drive cycle verification error (12.32 mV) is achieved, along with a high FI of 0.08 $V^2$, all within a short experiment duration of 1800 s. This reflects the ability of the RL-based OED to effectively excite the parameter \( k_n \) while being time-efficient. The RCID profile, while requiring a much longer experiment duration ($\approx$ 28 hrs), achieves the highest FI (0.42 $V^2$) and a similarly low verification error (13.33 mV), making it a robust but time-intensive approach. Traditional fixed profiles, such as 1C and 5C Discharge, exhibit significantly higher verification errors (47.98 mV and 47.31 mV, respectively) and lower FI values (0.01 $V^2$ and 0.05 $V^2$), demonstrating their limitations in sufficiently exciting \( k_n \), thereby producing data set that are insufficient for accurate parameter identification. These results underscore the RL-HIL's capability to deliver highly accurate parameter identification in a time-efficient manner while also showing that RCID remains a strong method for scenarios where experiment duration is less critical.

\section{Conclusion}
The importance of optimizing the input excitation to generating data sets for parameter identification is demonstrated by the favorable results of the RL agent's OED.  By maximizing the FI for a specific parameter over the duration of a 1800-second experiment, an optimal experiment design can be generated online by the reinforcement learning agent in real time. In a 30-minute experiment, the RL-based OED produces data containing high levels of parameter identifiability.  

This result is confirmed by utilizing the generated data set for parameter identification via minimization of the output error. The results indicate that the test profile generated by the RL-based OED leads to significantly lower voltage RMSE than conventional constant current discharge tests and is on par with the efficacy of a 28 hr RCID test.  By utilizing optimal experimental design, accurate parameter estimation can be achieved in a fraction of the time it would normally take to get values that result in such low errors. As an extension of this work, the proposed framework will be applied to generate OEDs for other model parameters or groups of parameters altogether.


\begin{ack}
The authors wish to acknowledge the Honda Research Institute for the inspiring discussions and helpful feedback that contributed to the research presented in this paper.
\end{ack}

\bibliography{ifacconf}             
                                                   







\end{document}